\documentclass{article}
\usepackage{amssymb}

\begin{document}

\title{Dialectical Roots for Interest Prohibition Theory}

\author{Jan Aldert Bergstra\\~\\
           \small{Section Theory of Computation, Informatics Institute,}\\
           \small{Faculty of Science, University of Amsterdam, Science Park~904,}\\
           \small{1098~XH Amsterdam, The Netherlands; email: \texttt{j.a.bergstra@uva.nl}}
          }
                 
\maketitle

\begin{abstract}
It is argued that arguments for strict prohibition of interests must be based
on the use of arguments from authority. This is carried out by first making a survey
of so-called dialectical roots for interest prohibition and then demonstrating that for
at least one important positive interest bearing financial product, the
savings account with interest, its prohibition
cannot be inferred from a match with any of these root cases.
\end{abstract}

\section{Introduction}
\label{sect-intro}
Interest Prohibition Theory (IP Theory) is concerned with the theoretical aspects of
interest prohibition. In~\cite{BM11a} a brief survey of IP Theory has been 
presented, together with a meta-framework which specifies its
scope and reasoning methods. The meta--framework is characterized 
as a so-called application specific informal logic (ASIL) named 
ASIL$_\mathrm{IPT}$. Interest prohibition has old roots in a variety
of ancient cultures. These roots take the form of objections, some of which are outdated
and some of which still have some or even significant
relevance. I will speak of dialectical roots because these roots take the form of
arguments with a dialectical (that is logical) form which may be 
sufficiently convincing to convey the justification of interest prohibition 
without reliance on external authorities.

The recent advance of Islamic Finance has brought
interest prohibition back to the agenda of financial research. The prominence
that interest prohibition has regained, calls for an investigation of its roots,
not so much in historic terms but in conceptual terms. That investigation may,
at least in principle, lead to the discovery of new dialectical roots, and as I 
will explain below, in effect it does.

\subsection{IP: simultaneously on the way out and on the way in}
Writings about interest prohibition from a conventional  perspective
invariably depict it as a proposition which has become outdated
because of improved economic insights and because of strengthened
legal thinking, often attributed to Jeremy Bentham, placing the freedom of contract
(including financial contracts)  in very high regard. The development of a 
permissive attitude towards interest went 
hand in hand with overcoming mistaken views on astronomy and biology
and with the development of restrictions on various forms
of human economic exploitation. In each of these cases Christianity, or in any case
powerful parts of Christianity, failed to
play a leading role in the modernization. As a consequence of these factors modern 
writing on interest prohibition from a conventional perspective is not informed
by any urge to view interest as a mechanism worth being avoided, 
prohibited, or prevented  to date.

Islamic Finance, however, which shows a positive momentum
since around 1970, propagates interest prohibition, and as a consequence IP 
is on the way up in some Islamic countries at least. 

\subsection{Objectives and results of this work}
In this paper I will survey known objections against the interest mechanism
and consider possibly novel objections against  interests.
I will assume that different objections may be grounded on different perspectives on money,
and consequently the issue of their consistency does not necessarily arise.%
\footnote{
Because different objections may pertain to different, and possibly inconsistent,
views on money.%
} 
It is not clear in advance that
any of the known objections against interest payment are still valid with respect to money 
in its current form. 

Writing this survey of objections against interest has been 
an independent objective of this paper, but I  have no doubt that
improvements of this listing, as well as of the descriptions therein can be found. The items in the
list will be referred to as dialectical roots, or dialectical root cases with the latter phrase
stressing the sometimes anecdotical form of the item descriptions. With the term dialectical, 
which I am using as an alternative  for the adjective ``logical''  and which I prefer because it
carries less formal connotations, I intend to express that these roots exist
in a setting where an assessment of their merits is in principle 
subject to a dialectical discourse (rational reasoning). This stands in opposition to the use of 
so-called arguments from authority. The demarcation between dialectical reasoning and 
argumentation from authority is not always clear. I do not intend to suggest that arguments from
authority must be weak, as if they are a cover-up for lacking conclusive arguments. There is
room for such arguments and and there is also room for a distinction between good, mediocre,
and bad arguments from authority.

The second purpose of this survey of root cases is technically more involved. 
I will consider a single
financial  product: a savings account
with interest (SAI). SAI was considered in detail in \cite{BM10b}, to which I refer for 
further information. Regarding SAI it is possible to draw a number of specific conclusions based
on an inspection of the inventory of dialectical root cases. Indeed I propose to believe
the validity of the following assertions:
\begin{enumerate}
\item Currently SAI is not permitted by mainstream Islamic Finance, this prohibition 
is absolute
in the sense that no positive interest rate returned by the bank to the lender is 
allowed. (Opinions differ about the question whether or not there are circumstances of 
extreme urgency in which the prohibition can be ignored.)
\item 
The prohibition of SAI  cannot be convincingly inferred from any of the dialectical 
roots that have been collected in the survey below.
\item 
The justification of the prohibition of SAI in any framework of Islamic Finance must
must be found in arguments from authority put forward by past and current  
Islamic scholars. These arguments from authority do not depend on any 
claimed similarity between one of the root cases for IP and SAI.
\item 
Because of the central importance of SAI in the conventional financial system, it is valid
to say that the justification of the methods of Islamic Finance currently depends to a 
significant extent on arguments from authority. 
\item
The dependence of Islamic Finance on arguments from authority, rather 
than it being exclusively based on the contemplation of dialectical roots is a 
non-trivial conclusion. (It constitutes 
the central conclusion of the paper.)  It may seem obvious
at first sight  that IP justification in Islamic Finance depends on arguments 
from authority but that is a matter of appearance rather than a matter of fact in the light of 
the following observations:
	\begin{itemize}
	\item  
	The Church opposed to interest, after the middle ages, on the basis of 
	Aquinas' reasoning which is of an intellectual kind rather than of a religious kind. In 
	other words Aquinas' dialectical root(s) for IP has undone IP from its religious
	character (at least in the context of the Roman Catholic Church), and 
	as a consequence it has taken IP outside the area dominated by 
	arguments from authority.
	\item
	Many texts on Islamic Finance that explain IP as being rooted in arguments from 
	authority do not bother to argue that no alternative (and thus less religious) 
	arguments can be established for IP (compliant with an Islamic context of course). 
	\item
	Maximizing rather than 
	minimizing reliance on arguments from authority seems to motivate much 
	writing on Islamic Finance.
	\item Only by taking a single financial product (in this case SAI) as an exemplary case,
	the demonstration of the unavoidable role of arguments from authority can be 
	made to work.
	\end{itemize}
\item
Obtaining a convincing application of arguments from authority to the particular case of
SAI requires a meta-theory and more specifically a structure theory of arguments from authority
in the Islamic tradition. Such a meta-theory seems not to exist at this moment.
\end{enumerate}

\subsection{Why does this all matter?}
Developing a meta-theory for arguments from authority applicable in the setting of 
contemporary mainstream Islamic 
Finance is not an objective of this paper. The rather modest  point made here is to
demonstrate that such an analysis is indispensable for understanding IP as occurs in
Islamic Finance. 

The relevance of this particular conclusion transpires when one looks through 
introductions to Islamic
Finance, where quite often some argument against the interest mechanism
is suggested (invariably connected with
one or more of the dialectical root cases listed below) as being a plausible 
argument for IP.%
\footnote{See for instance \cite{TurkAriss2010} for an example of such an analysis.%
}
 Such arguments
are all misguided. None of these arguments can explain why SAI must be 
categorically prohibited, and not merely preferentially 
avoided or simply have its interest rate minimized. The remarkable reward which 
authors obtain from the use of 
arguments based on one or more of the dialectical roots is that they need not be 
clear about the arguments from authority involved in the matter. If an appeal is made
to an argument from authority that appeal is usually phrased in quite simplistic terms. 
For instance by asserting that ``it is well-known that Islam prohibits all 
forms of interest''. For the reader it is an almost
impossible task to find out why that assertion
must be confirmed.%
\footnote{
Scholarly authors contemplating the assertion that ``Islam prohibits all forms 
of interest'',  all seem to dispute a strict interpretation of that 
somewhat uncompromising claim.
For instance see \cite{Far05a,Far07a} and \cite{Rahman1964}. These and other 
scholarly works on the subject assert that Islamic Finance proponents' claim of interest prohibition
factorizes into two independent separate claims: (i) {\em riba} is forbidden and, 
(ii) the so-called equivalence view: {\em riba} equates interest. 

For (i) I refer to \cite{Nomani2002} which unambiguously demonstrates the complexity of 
understanding that claim. For (ii) one may consult \cite{Far05a} and many other sources,
all testimony to the difficulties involved in establishing the equivalence view.%
}

\subsection{Preliminary remarks concerning interest}
As a preparation for the real work of the paper I provide some comments on (i) the terminology
about interest, (ii) the selection of SAI as a special interest bearing financial product which
can play a pivotal role in the demonstration that arguments from authority must play a central
role in Islamic Finance, at least as long as its rejects SAI, and (iii) the role of IP in the characterization of Islamic Finance. 

\subsubsection{Usury versus interest}
Following \cite{Ackerman1981} I will assume that interest prohibition has throughout 
history emerged (and re-emerged) after
phases in which interests are common practice. Philosophical and theological considerations,
sometimes quite remote from the economic theory and practice of the day, have 
mostly driven the moves towards interest prohibition.  Thus IP is  (and always has been) a reactive phenomenon regularly gaining momentum after phases where IP has been
nearly forgotten. In \cite{AliKhan1929} it is reported that quite in the Islamic world
strict rules for IP were often evaded by high ranking authorities
 when the economy of the state was at stake.

It seems impossible to contemplate IP in the context of Islam without entering
in a complex debate concerning the historical meaning of technical terms
as expressed in terms of
their somehow well-known contemporary relatives. 
As an illustration of this situation I consider the well-known
contrast between usury and interest about which \cite{EkelundHeTo1989} asserts the 
following (rendered in
my own words), concerning the meaning of these terms in Europe in medieval times:
\begin{description}
\item{\em Usury.} Usury derives from the Latin ``usura''. It denotes the reward that a borrower 
of a loan returns to the lender (of that loan) justified by the profit the lender makes on 
operations enabled by the borrowers control over the means constituting the loan.
\item{\em Interest.} Deriving from the Latin ``interesse'' denotes the compensation
made by the borrower for the the lender's opportunity costs implied in providing the loan.
\item{\em Risk.} Risk of borrower default was ignored because borrowers were normally
assumed to be in possession of significant real estate of a value far in excess of the loan.
\end{description}
This explanation may be contrasted with common assertions that either (i) usury is an 
old term for
interest but with the same meaning in essence, or that (ii) usury refers to interests which 
are in some sense excessive and are for that reason to be considered especially problematic.

Below I will take interest to be a combination of usury and interest as given above, possibly
including a compensation made by the borrower to the lender  for borrower default risk.

\subsubsection{SAI: the simplest case of a financial product involving interest}
In~\cite{BM10b} the point of view is taken that interest prohibition implies a reduction
of the set of permitted financial products. ``Reduced Product Set Finance'' was coined
as a phrase to indicate a financial system grounded on financial product restrictions. 

The clue that a system may potentially be strengthened by removing some of its most
useful features has a long history in computer science: RISC (Reduced Instruction 
Set Computing) gained an advantage by simplifying processor logic in favor of 
processor speed. Functional programming removes the assignment statement, in spite of its
central role in all imperative program notations; scripting languages often ignore typing of 
data. In these cases a seeming defect is turned into a (claimed) advantage for the resulting programming methodology.

A search was performed in~\cite{BM10b} to determine the simplest 
conceivable financial product
which becomes forbidden under a comprehensive exclusion of interest payment.
It was concluded that this simplest case is found with an interest bearing savings 
account hosted by a bank offering maximal protection (in comparison with
competing financial institutions) against default risk. 
Upon further inspection that mechanism turns out to be quite complex and
the question what exactly is supposed to be forbidden becomes harder to grasp when
the financial product at hand is specified in more detail.%
\footnote{%
Without further elaboration I mention that for an adequate understanding of SAI it is
important to imagine that the borrower will be providing that service to several 
clients in parallel. A formalization of concurrent systems which has sufficient expressive
power to model that form of parallel activity is the thread algebra of \cite{BM07c}.%
}

Below I will write as if SAI is self-explanatory; my belief in the pivotal 
role of SAI in the issues concerning IP stems from the following observations.
\begin{itemize}
\item As interest rates decrease the substance of injustice that SAI may create decreases 
as well. Accordingly no argument based on expected injustice can explain that SAI must be
forbidden for all (especially small) interest rates. It is the categorical 
nature of IP in this particular case that calls for attention.
\item One may say: first prove that Islam prohibits all forms of interest, and subsequently 
apply that
result to SAI. Here I propose that the intuition of cut-elimination in proof theory is not
omitted too soon. If this kind of proof exists then a shorter proof (that IP applies to SAI) must
exist which makes no use of the outermost application of modus ponens.
\item It may be assumed that the borrower (the bank, guaranteed by the state) will not default.
Thus a complete separation of compensation for the lender's opportunity cost from the compensation from the lender's exposure to borrower's default risk is obtained. This is the
significant simplification that SAI provides with respect to financial products for
consumer credit. 
\end{itemize}

\subsubsection{Islamic Finance without interest prohibition}
One may think that Islamic Finance cannot be separated from interest prohibition,
so that the problem if and how IP can be justified
concerns the rationality of Islamic Finance
straightaway. This issue has been raised in \cite{BM11b} from which I take the following 
analysis.

Questioning the arguments for IP is not equivalent to questioning the rationale of 
Islamic Finance, because, remarkably, the rationale of today's Islamic finance 
becomes conceptually simpler if
no mention of interest prohibition is made at all. Although speaking of
interest prohibition provides a fast track to an explanation of the current structure
of Islamic finance, that fast track is found at the cost of leaving the questions mentioned in
\ref{questions} below in need of an answer.  

A simpler explanation of Islamic Finance, which is
not based on an axiomatic assumption of IP
for, either in general or for any particular financial product, is that Islamic Finance
consists of the theory and practice of using a growing portfolio of financial products
which are considered flawless by a  large majority of Islamic scholars.%
\footnote{
Under this
interpretation the significance of interest is merely that upon its detection (in 
a financial product specification, however marginal the occurrence may be
at first sight) the 
opposition of a significant number of Islamic scholars can be predicted. The essential
step thus proposed  is to move interest prohibition from a principled status to the status of a
heuristic instrument usable for Islamic financial engineering. The precise form of this
heuristics may, however, fluctuate in time, unlike the status of generally 
approved and centuries old priniciples.%
}
Having this convention in mind Islamic Finance need not disallow any financial products
on an indefinite basis. By being strict when admitting a financial product the creep of
unwanted practice is stopped, while a more liberal standpoint in philosophical terms
(if ever obtained) can be implemented by simply admitting a larger product family.

Now this picture may be too simple as well. Tarek El Diwany,
in~\cite{ElD97a}, which I have surveyed and commented 
upon in~\cite{Ber11b} remarks that although the modern product line of 
Islamic Finance has obtained wide-spread scholarly approval, its realization 
in Islamic countries 
seems to have had side effects that must raise doubts about the plausibility of 
today's mainstream approach to Islamic Finance.

\subsection{A wider context}
The conclusions of this work should preferably 
be assessed in a wider context than the limited theme of IP justification. 
An attempt to grasp a wider perspective leads me to three 
separate observations: (i) it will be worthwhile to investigate chained arguments from authority
because these are likely to have an interesting structure, (ii) worries that have been 
voiced about the justification of IP in the setting of Islam: three such worries are phrased
as questions and as it turns out the final dialectical root case provides a context where
these questions have an answer that takes care of the mentioned worries, (iii) the project
of developing an ASIL for IP is an ongoing one, and some implications of this work for the 
design of an ASIl in this context are collected.
\subsubsection{Remarks on arguments from authority}
I hold that general assertions like ``Islam prohibits all 
forms of interest'', presented without any further justification, have the opposite effect of what
may be intended by an author who makes that assertion: to diminish the strength of 
the argument from authority at stake rather than to enforce it. It
makes one's belief in the justification of IP decrease rather than increase.
By being sloppy about arguments from authority an author contributes to their erosion,
rather than to their enforcement.

Being precise about lengthy arguments from authority is a complex matter which seems to 
have been entirely neglected by professional logicians as if there were no conceivable
need for it.
Without going into details I propose
that at least seven levels of authority are involved in the reasoning chain that constructs a 
justification for the prohibition of SAI:
\begin{enumerate}
\item The authority of God's judgements  pertaining to matters of finance.
\item The authority of the prophet's phrasing of God's views pertaining to matters 
of finance. (I refer to \cite{Don10a} for a description of the context of the origination of these
phrasings.)
\item The authority of the wording chosen by those who have put the prophet's words on paper.
\item The authority of the Islamic scholars (second and third century) who generalized the prescriptions found
in the Qur'an pertaining to IP, to a more universal claim of IP, then put forward as a authorized reconstruction of
God's will in the matter from the revealed sources at hand.%
\footnote{In~\cite{BM11a}
it is outlined, following~\cite{Sub01a,Far07a} that the Qur'anic prohibition of
interest made explicit mention of the so-called doubling debt scenario
and that subsequent authors (for instance El Jassar, see~\cite{Far07a}) have 
successfully and authoritatively argued
for an extension of the principle of interest prohibition to many more scenarios
of coupled lending and borrowing. Their arguments derive from the general 
methodology of Islamic legal and ethical theory
construction, in particular (i) that original sources may be understood
as having been stated in an exemplary form, so that more general truths lying 
behind these particular cases can and must be uncovered by subsequent
intellectual work, and (ii) that older subsequent work is considered to be
more authoritative than later work.%
}
\item The authority of subsequent Islamic scholars who indicated by what priority mechanism
the variety of primary and secondary sources of Islam, with relevance to IP, is best understood.

\item The authority of current Islamic scholars who state that IP holds for all (or at least some, including SAI)
interest bearing financial products which have been developed for today's money.
\item The authority of financial experts who claim that IP is not only wanted but that it is also practical to such an
extent that maintaining its prohibition (including the prohibition of SAI)  is an economically coherent position.
\end{enumerate}

This listing only serves to indicate the unavoidable complexity of arguments from 
authority that on encounters if IP in the special case of SAI must be motivated.%
\footnote{Assuming the analysis is made in English an additional problem requires
careful attention.
For  any analysis of arguments from authority concerning IP the role of the 
Arabic language presents a very serious problem. 
Questions about the scope of interest prohibition, the 
rationale of it, its history and its ethical basis are often confused with issues about the 
meaning of Arabic terms, the history of that meaning and the particular use made
by authors of these terms. In~\cite{Ber11a} it is suggested that the entire discussion 
can be performed in English, (or whatever non-Arabic language indeed), assuming 
that translations of a number of Arabic terms are available. 

Satisfactory translations of important terms seem not to exist and it may take years before 
workable translations stabilize. Perhaps there will be a need for different translation 
strategies that incorporate specific but mutually incompatible viewpoints throughout the translation of a range of terms.%
}
An appreciation of the complexity is required for understanding the relevance of the
conclusion, to which most of this paper is devoted, that arguments from authority are
essential for the justification of IP in an Islamic context.

\subsubsection{Major conceptual questions, accessible to non-Muslims}\label{questions}
To see the depth of the problem that an Islamic directive of interest
prohibition presents, in particular when based on arguments from authority, 
I make use of A. Subhani's discussion in \cite{Sub01a} of that issue: 
Given a chain of arguments from authority against SAI, still these problems remain:
\begin{enumerate}
\item
Why is punishment for offenses against interest prohibition  limited to so-called
ecclesiastical%
\footnote{Ecclesiastial punishments is the term used by Subhani in~\cite{Sub01a}. The 
explanation indicates that these are punishments to be issued by God after one's 
death, instead of being administered by human courts and authorities, during the
trespasser's life.}
punishments. 
\item 
Why are the penalties promised to those who fail to comply with interest prohibition so
excessively severe, which is made more puzzling because intuitively the failure becomes
less when the amounts of values involved decrease.%
\footnote{In particular if so-called concealed interest (as a phenomenon implicit in
the use of some particular product) is uncovered after the product has been
designed and after a phase of initial usage,
it is incomprehensible that someone be severely punished for not having seen 
that fact before making use of the product. Detecting the occurrence
of concealed interest has become
an area of specialization inaccessible to average citizens.%
}
\item
Why are interest bearing financial transactions in which lender and borrower 
operate freely to a well-understood, stable, and mutual advantage, while
not doing any conceivable harm to any third party, forbidden at all?
\end{enumerate}

As  conceptual problems these questions are  perfectly accessible to a non-Islamic person. 
The matter is intriguing by all means, irrespective of one's religious position. The fact that the
imperative against interest payments predates the origins of Islam more than
1000 years only adds to the weight of the issue: what was the problem that so many 
people claim to have seen?

\subsubsection{Consequences for an ASIL for IP}\label{CR}
The development of an ASIL (Application Specific Informal Logic) 
for (Islamic style) IP, which has been proposed in \cite{BM11a}
is an ongoing process. From the work in this paper some notions can be extracted which
are of the generality that these merit a place in the ASIL. This includes the following matters:
\begin{enumerate}
\item An ASIL for IP must contain a structure theory for arguments from authority, which is
sufficiently flexible to allow a uniform presentations of different chains of arguments and to allow
the comparison of such arguments in terms of their force.
\item It seems
reasonable to assume that a logic for inference by means of authority can do without general 
universal assertions which are first derived and subsequently instantiated. Proving that fact
about arguments from authority requires
what is called cut-elimination in logic and it may be formulated as a requirement on the ASIL for IP that it will allow an analysis of the power of cut-free proofs.%
\footnote{%
The argument that cut-free proofs may be expected for ethical issues is not self-evident. For
instance the assertion that person P is entitled to protection from body B
may be obtained by first deriving that all human beings are entitled to a certain degree of protection by B. In this example deciding that A is a human being is assumed to 
be unproblematic. In the case of interest generating financial products, however, it 
may be quite difficult to decide about the presence of interest. Thus a universal
quantification over all forms of interest is semantically more problematic 
than a quantification over 
all human beings, assuming that physically problematic cases are ignored.%
}
\item The contrast between arguing from dialectical roots and arguing from authority requires
a firm footing: different inference rules seem to be at work.
\item Dialectical roots can be used in an argument 
against a financial product if some similarity between the root case and a product instance can be established. ASIL for IP should provide guidelines for
proving and disproving these similarities. Below I will argue that none of the dialectical root cases
offers a similarity to SAI which by itself induces a prohibition of SAI assuming that one
shares the injunction embodied in that particular root case. The weaknesses of this line of thought must be admitted: (a) there is no analysis of the combined implications of two or more root cases, and (b) there is no analysis of the degree of dissimilarity that must be established before
a root case can be correctly claimed to be irrelevant for the product at hand 
(SAI in this particular case).
\item Adoption of the following continuity rule: 

if an argument against a financial product involving a positive  
(that is non-negative and nonzero) rate of interest 
depends on expected negative or unjust consequences of its adoption or use, then for some positive interest rate these
negative consequences must be negligible (continuity of cause effect relation). Thus, such
arguments cannot prove that all positive interest rates (for that specific product) are to be 
rejected.
\end{enumerate}

\section{Generalized phrasing of the interest mechanism}
Although interest is always connected with money, the mechanism which has become
criticized throughout history is does not require money and not even the existence 
of money. Below the intuition at stake will be captured in terms of critically productive transactions.
\subsection{Critically productive transactions}
From~\cite{BM11a} I include the notion and specification of critically
productive specifications, further distinguished in lender side and borrower 
side critically productive transactions.

\begin{quote}
We say that a transaction between parties A and B is Lender-side
Critically Productive (LCP) if A acquires goods, services or valuables
G in compensation of lending valuables V to B for some limited duration.
It is assumed that these valuables serve no instrumental purpose for B
or any other agent other than serving as a means of exchange or as a
store of value.
We note that:
\begin{enumerate}
\item
the generation of G is the productive aspect, more specifically B is productive;
\item
V may be understood as an asset which represents wealth;
\item
B or B's sources will not use V as a tool of some kind during the period of the loan;
\item
the apparent circumstance that A is not involved in any substantial
fashion in the generation of G constitutes the critical aspect of the
transaction.
\end{enumerate}

In a borrower side Critically Productive transaction (BCP transaction), 
the borrower of valuables produces assets V
without sharing these with the owner of the valuables in any pre-agreed
fashion.
\end{quote}

Any financial product that features interest or concealed interest has a part
of its behavior that matches with an instance of BCP or LCP. Instances of BCP and of LCP
are found
by first instantiating the asset class parameter for valuables, the asset class
parameter for goods and the process parameter for the protocol
(transfer scenario). The result of such an instantiation is still generic and it must be 
instantiated further to match with (a component of) a specific 
financial product.

\subsection{Parameters for Critically Productive Transactions.}
The mechanisms of LCP and BCP transactions depend on several
underlying assumptions of which the most important ones are listed here.
\begin{description}
\item{\em Asset class specification for valuables.} 
Conventional interest prohibition
is obtained when LCP is instantiated with money as the asset class for valuables.
I will distinguish the following asset classes:
	\begin{description}
	\item{\em	Exclusive valuables:} 
	hard to get, stable in value, storable, durable, protectable, 
	appealing, symbolic, non-personal, objective, carrier of wealth (for instance:
	gold, silver, very fine works of art), usable as a means of 
	exchange and as a store of value.
	\item{\em Non deprecating commodity money:}
	ideal money: a hypothetical category? Just like the exclusive valuables
	but not exclusive and not primarily a sign of wealth, but merely a 
	proof of purchasing power.
	\item{\em Investment certificates:}
	including shares, bonds, participations to equity funds; assuming that revenues
	are  re-invested in the same vehicle.
	\item{\em Fiat money:}
	including bank money.
	\item{\em Intellectual property rights:}
	including patents, copy-rights, design rights, licenses,
	\item{\em Political power:}
	correctly obtained power in a reasonable political system.\\
	\end{description}
\item{\em Asset class specification for produced goods.} 
Nowadays everything is 
expressed in money, but that is not to be taken for granted in a root case
for a proposed conjectural history.

Each of the mentioned asset classes may be prescribed for a particular
case of a loan. Moreover the following asset classes may be appear
for containing goods returned for obtaining a loan (or earned by making use of
a loan):
	\begin{description}
	\item{\em Non-consumable goods:}
	real estate, furniture, tools, weapons, clothes,
	\item{\em Consumables:}
	including: food and drinks, cattle, oil, wood.\\
	\end{description}

\item{\em Transfer scenario.} The protocol of exchanges of promises, agreements,
payments and claims. Different scenarios lead to different prohibitions. Different
phases of a scenario may prescribe transfers in terms of different asset classes.

I will make use of the straightforward transaction patters only where all steps
are specified in advance and no decisions on how to proceed are taken in 
between except for the possibility that the borrower must tell the lender at
maturity dat that he will not be able to return the principal amount.
\end{description}

CPT's (critically productive transactions) as specified above has two parameters
V for an asset class for valuables and G for a more extensive asset class of goods, 
including valuables.
CPT(V,G), (that is CPT's with valuables V and goods G) is still very general and it
can be instantiated in many ways, for instance: (i) CPT(EV,G) denotes
critically productive transactions with exclusive valuables (EV) as valuables and 
any goods
as goods. (ii) CPT(NDCM,NDCM) has valuables and goods both in the form of 
non-deprecating commodity money (NDCM). (iii) CPT(FM,FM) is the current 
situation where 
both valuables and returned goods are given in terms of fiat money (FM).

Nowadays the most frequently occurring type of CPT is CPT(FM,FM), which
has many further realizations such as for instance a savings account and
a consumer credit.

\section{Dialectical roots for IP I:  a survey of old roots}

Below I will outline the known dialectical roots of interest prohibition. 
The listing is  exhaustive to the best of my knowledge. Cases that are
primarily in favor of interest moderation are omitted, because such cases don't touch  
the fundamental
questions concerning strict interest prohibition and also because  non-Islamic finance 
acknowledges the importance of moderation so
that views from different sides practically coincide on that matter.
Laws against excessive interests rates, but permissive of acceptable rates, and usually 
issuing upper bounds only, have emerged in many phases of economic history.
Each individual case is best understood together with a specific perspective on money,
which supports its plausibility.%
\footnote{In \cite{Ber10a} I have made an attempt to provide a survey 
of different perspectives on money.%
}
These cases may serve as roots for a 
focus on interest prohibition, 
while not committing to consequences which may have obtained a wide audience
but which may not have been established beyond all reasonable doubt.

\subsection{Judaic tradition: don't charge interest from your own people}
Perhaps the oldest objection against interest payment suggests that one should not
ask it from one's own people (or may I say tribe). In the same tradition debts should be
outstanding for a limited period only. 

This restriction makes most sense if there is no inflation. Inflation has the effect of 
resolving debts on the long run, thus mitigating the problems caused by debts.

It seems to be the case that in the Judaic tradition the borrower was considered to be
the weaker party by definition. In addition interest free loans were preferred over
gifts because a gift may induce a status problem for the receiver.

\subsection{Aristotle: accumulation of money for its own sake is wrong}
Aristotle argues that money is an instrument that helps trade. Using trade as a method
to increase one's stock of money is an inadequate use of the tool and so is 
lending money in order to receive an interest. Both scenarios represent an unnatural
usage of money and are wrong for that reason, the interest harvesting scenario being
the worst of the two as it is not even based on proper transactions.

This restriction makes most sense if the money and its circulation has been designed 
in such a way that each agent has an empty stock of money on regular moments, just like the 
fuel tank of a car. The restriction seems not to pertain to the prohibition of interest rates
that are fixed in advance. Rather it outlaws any transaction exclusively intended for 
monetary advance.

\subsection{Qur'an: doubling debt scenario (DDS) not permitted}
The doubling debt scenario involves a debt which is doubled if the
borrower  must admit that he will not repay the principal sum at the agreed
time of maturity. There are variations on this theme with and without the agreement
to pay an interest to the lender with regard to the first period, and in the most
liberal interpretation of interest prohibition only an exponential growth by
way of repeated redoubling of that interest sum proper is prohibited.

DDS suggests an extreme case of excessive interests but it all depends on the length 
of the period, with nowadays common interest rates below 10\% it may take no 
some 10 years to reach duplication. So if the period in DDS is 15 years that
corresponds to an interest rate which would not be considered excessive today.

The DDS prohibition involves an interest rate (100\% per period), and it also involves
making a plan or protocol about transfers between lender and borrower 
in advance. These elements have been taken on board by
all subsequent authors opposing the interest mechanism.%
\footnote{
The doubling debt scenario (DDS) may be considered only an example from a wider
range of illegal practices, which was explicitly mentioned explicitly in
the Qur'an perhaps because it was applied frequently in those days. (But a proof of that
historic basis for rejecting DDS seems not to exist.) 

Around 1000 AD, using abduction, a more general imperative  
has been developed, which is nowadays understood as a universal objection 
against the use of interests in any conceivable case.
This generalization impacts on today's design of the imperatives of
Islamic finance. It fails to constitute a dialectical root case, however, as it 
primarily applies
an argument of authority to the methodology of reading a specific collection of
revealed sources.%
}

\subsection{Medieval Roman Catholic Church: savings monopoly}
By prohibiting interest paid for consumer credit the state can make it more
attractive for citizens and organizations to place deposits originating from
savings under state controlled institutions, thus helping to finance the state. 

In \cite{EkelundHeTo1989} the hypothesis is formulated that this mechanism has 
motivated the Church's ongoing endorsement of interest prohibition, although it is
not considered explanatory for the origination of that endorsement.

\subsection{Aquinas: double compensation objection}
An important explanation of the rationale of interest prohibition is found in 
Aquinas'  treatise ``De Malo'', which has probably been written in Paris
between 1270 and 1272, on the basis of preliminary work done
between 1265 and 1268 in Rome. A recent and very readable translation to 
German is given by C. Sch\"{a}fer
in \cite{Sch10} (Questio 13, {\em Artikel} 4, pp. 152-165). Many arguments in 
favor of the payment of interests are known to Aquinas, including the concept of opportunity 
costs, but nevertheless he decides that it is strictly forbidden ({\em Tods\"{u}nde} in 
Sch\"{a}fer's translation). The key 
argument is two-fold: (i) All items when used are either left as they were, perhaps with
minor damage or wear (for instance a house that is rented) or are actually consumed
(such as water or food). Money is of the second category, as its use is destructive at least
from its user's perspective. Thus its use cannot be sold, only the good (that is money)
itself can be transferred. Asking for an interest is asking for double compensation: for the 
principal sum proper and for the usage thereof but these two forms of usage
are exclusive. (ii) The claim
of interest in connection with lending is unnatural, and for that reason it is forbidden. 

The biblical texts mentioned by Aquinas are all inconclusive and for that reason he uses
a style of reasoning dating in part back to works of Aristotle. Aquinas finds it permissible to lend
money against a zero interest rate, under certain circumstances that is even considered
a virtue. 
Remarkably interest may be charged according to Aquinas if a promise is made by
the borrower that the money will not be used, at least not in its monetary capacity. Here
is an example that I designed using some ingredients from the text of Aquinas: a
lender A may lend golden coins to B, which the borrower melts and then turns into a 
golden bowl, for the use of which he will pay A some amount (to be understood as a 
rent rather than an interest). At maturity of the loan B will transform the bowl back into the 
original volume of coins and redeem the amount due to A.

Aquinas failed to appreciate the circulating nature of money, at least in the mentioned text;
 its use may
seem consumptive or destructive, one time only, like that of food, at first sight, 
but on the long run the money returns. It is not easy to see why Aquinas considered
his arguments to have been conclusive himself, except for the fact that he may have felt 
somehow obliged to steer towards the conclusion he gave in his answer.

Sch\"{a}fer states that Aquinas' intention with the entire book was not to primarily
to promote his own points of view, but rather to collect a range of existing sources in a 
unified manner.

In Summa Theologica, Question 78, ``the sin of usury'' (see \cite{NewAdvent1920}) Aquinas 
presents less comprehensive but similar arguments, though now concluding that borrowing 
with usury from a professional lender used to accept usury, may be permissible if it is done 
with good purpose. In the Summa Biblical references against usury are not deemed
conclusive and like in ``De Male'' technical
arguments against usury are put forward as being decisive.

\subsection{Subhani: ex-sui creation trespasses on God's territory}
From A. Subhani's contribution ``Whither Islamic Finance...'', as posted on New Horizon 
(January 2011) I take this quote
which pinpoints ex-sui creation as a right of God which man should not reserve 
for himself. Avoiding ex-sui creation
is a motive for opposing to interest bearing loans.
\begin{quote}
To sum up, the operative principles of Islamic finance require that the internal 
process of the transaction, which is the focus of Islamic law and not the effect 
of the transaction, must incorporate cogeneration and eliminate self generation, 
delay (except under dire necessity) and risk-taking.
These operative principles of Islamic finance are consistent with the theoretical 
model of creation, which conceptually consists of (i) creation from nothing 
(ex-nihilo creation); (ii) creation from the object itself (ex-sui creation, to 
coin a new term) and (iii) creation from other object (ex-alio creation, to 
coin yet another new term). Ex-nihilo creation is expressly and exclusively 
a divine capability and hence any human action even resembling ex-nihilo 
creation (e.g. delay in transaction settlement, as explained above) stands prohibited. 
Similarly, ex-sui creation is expressly and exclusively a divine capability and 
hence any human action even resembling ex-sui creation (e.g. interest on money) 
stands prohibited. These two prohibitions constitute the essence of the prohibition 
of riba. Ex-alio creation alone is a human capability and hence any human 
action that incorporates ex-alio creation (e.g. murabaha in the commercial 
domain) stands permitted. This permission is the essence of the permission 
of bayÕ. Finally, with the Islamic legal focus on generation (the prohibited 
self generation and ex-nihilo generation and the permitted co-generation), 
speculation (risk-taking) automatically stands prohibited, yielding a 
cohesive theory of Islamic financial law, at once monotheist, moral and ethical.
\end{quote}
The argument against ex-sui creation is a defeasible argument. It can 
best be imagined if money is gold or silver. But if, for instance, the money is gold then it assumes
in addition the silent assumption that the prospective borrower is not in the possession
of a gold mine which he can exploit after receiving an initial loan from a lender, but which
he cannot exploit without a loan or a third party investment.

\section{Dialectical roots for IP II:  new roots}
Besides the old root cases other principled arguments against interest can be imagined.
Below some listing of those is given. Completeness cannot of be guaranteed of course
(defining its completeness is problematic too).

\subsection{Strictly bifunctional money}
Money is often characterized as a being a means of exchange  (MOE) and constituting a 
store of value (SOV).
If one designs a financial system in which money serves these two roles only 
(strictly bifunctional money), then that
system will not generate interest payments, because it serves neither of these functions.

In abstract data type theory an initial algebra exclusively satisfies the equations constituting
it specification. I(MOE+SOV) may represent an ``initial'' design of money specified by MOE
plus SOV. The virtual omnipresence of the interest mechanism suggests that  either 
a financial system compliant with the specification
the name I(MOE+SOV) does not exist, or it exists and in addition it provides for interest payments.  

\subsection{Circuit theory}
Circuit theory proposes to prove that widespread lending against interest can only 
be done if either the amount of money increases with inflation as a result of
if the economic activity grows so that the amount of money can grow without causing inflation.
Assuming a stable economy as the intended equilibrium state rather than a growing economy,
the observations made by circulation theorists indicate interests as a cause of inflation, and 
for that reason as a mechanism that must be avoided.

Indeed by lending money to a collective of borrowers these are placed in an impossible position
because the interest that they have to pay are physically nonexistent, at the time the loans
are initiated.

If one assumes that a fraction of the loans will not be repaid due to borrower default
this inconsistency disappears, however.

\subsection{Interest prohibition for ideal and fair money.}
Ideal money is fairly accessible to anyone, it is insensitive to inflation or
other forms of deprecation. A plausible way for it to have been distributed is
by a state agency making sure that everyone gets sufficient means of existence.
Now the simple principle is that ideal money should only be used by its owner,
although technically it can be as well used by anyone else.
Lending it does not change ownership. It follows that offering a compensation 
for lending or obtaining an advantage out of borrowing is simply not plausible,
as long as the tight coupling between owner and user is maintained.

\subsection{Personalized electronic money: access to one's own money only}
If, in a near future all money has become electronic, each occurrence of it may be
tagged with its owner's identity. During a transaction the owner is modified. Each 
transaction must be ``real''. A credit sale of money can be forbidden. Now lending
becomes impossible (or at least useless), if it is additionally required that spending
money can only be done by its owner.

Strictly speaking this is a technical argument against debt. In these circumstances
borrowed money cannot serve as a manifestation of purchasing power (MOPP) because it
cannot be used by the borrower as a means of exchange (MOE). As a consequence there
is no basis for either asking or paying interests. In fact it becomes plausible in these circumstances that interests are negative and compensate the borrower for the cost of
holding the loan until the lender is able to take responsibility for the principal amount 
again.
\subsection{Limited double compensation objection}
If the double compensation objection is take more strictly, interest can be paid by the 
borrower
for that fraction of a loan which is not used for expenditures. More precisely
if a loan has the effect that on a number of days the borrower has an additional 
positive amount of money available, he may pay a rent for that money (renting it
only when and to the extent that it constitutes a part of the borrowers positive balance) 
because he is using it not ``as money for purchasing objectives'' but only to feel 
secure about his purchasing capability. (This seems to be consistent with Aquinas' views.)

Assuming that the borrower's spending is not known beforehand, the precise amount due
as rent for the principal sum (computed along the lines just mentioned), is only known at
the date of maturity of the loan.

I will assume that spending is done on a FIFO (first in first out)  basis, if more loans are obtained. 
Indeed
the situation gets more complex if a borrower obtains different loans at different moments. 
Interestingly later loans will probably produce higher returns to the lender because they
add more to the integral of purchasing capability.

In this view the money cannot be abstracted from its physical presence: FIFO requires
a time stamping of monetary tokens.

Taking Aquinas' analysis for true, there must be a permissible rate for borrowing money 
money that is not used for expenditures during the period that it is lend. Let this 
rate be $p$, 
then one might say that a double amount can be borrowed at that rate under the 
condition that half of it will be kept in store by the borrower. This proves that what Aquinas
writes is permissive of interest in the case of SAI rather than dismissive, assuming that some constraints on usage may be imposed.

In terms of a perspective on money one may think of I(MOE+SOV+MOPP), money that 
exclusively serves as a means of exchange (MOE), a store of value (SOV), and a 
manifestation of purchasing power (MOPP). It is the only latter role that qualifies for an admissible compensation by the borrower.

\subsection{Credit money only: service replaces interest}
Once cash has disappeared and all money has become electronic it is conceivable that
all money used by consumers is debt to a bank. Consumers may choose a bank where they
intend their incomes to be accumulated through transfers from other banks. In this situation the 
non-institutional agent cannot have free floating money that he does not lend to anyone.
The only choice available to the client is the selection of an institutional borrower. Now clients
may not wish to perform that selection on the basis of interest rates.

It is unsatisfactory that banks compete for clients by offering high interest rates which at the same
time undermine a bank's financial stability.%
\footnote{The experiences made with the so-called Icesave problem that hit savings 
account holders from the UK and the Netherlands
as a consequence of the 2008/9 financial crisis, confirms that this issue is real. 
The protection
of consumer property against a malfunctioning interest driven competition for consumer savings
has proven defective and the legal resolution of the matter has proven to be difficult. 

Focusing on the Dutch situation the following can be added: astonishingly savers who opted 
for putting their savings on an Icesave account on the basis of promised 
high interest rates were afterwards portrayed as having been greedy and having been
irresponsibly unaware of the fact that they had to take
many more factors into account when making decisions concerning savings account 
selection and management. At the same time Dutch
authorities (DNB, De Nederlandse Bank) felt unauthorized to inform the public of the 
particular risks that were known to them already. Indeed these risks may get out of control once their existence is made public by DNB, to the extent that, paradoxically,
communicating 
about the risk becomes by itself a very problematic risk factor. 

One may compare this to a fire-alarm which is left
unused because of the risk that it induces a panic. Or with a vital error signal that is not flagged with in an airplane in order not to deflect the pilot's attention. For some reason
financial risks can hardly be decoupled from the euphoria or panic that surrounds them.

Local municipalities were supposed to maximize expected income from
interest rates when placing their savings (while risky investments had already been labeled as
being inappropriate for tax generated money)  and municipalities were not even permitted to 
investigate the stability of the different banks, that
investigation being the sole task of financial authorities, who, however, kept silent instead of 
informing the public (including said municipalities).%
}
Instead of attracting customers by promising a relatively 
high interest rate on their credit accounts a financial institution may offer other services and 
advantages, possibly including interest free credits for particular purposes, or free medical service
under certain conditions, in order to satisfy a broad spectrum of (prospective) client needs.

\subsection{Prohibition of CPT(EV,G)}
I will now use the formal notation for transactions (that is financial products) of
critical productivity type. This allows to formulate a reasonable ``new root'' for 
interest prohibition.

CPT(EV,G) prohibition expresses the prohibition of both LCP(EV,G) transactions 
and BCP(EV,G) transactions.%
\footnote{
To illustrate the problem caused by violation of CPT(EV,G) I will consider an
analogical situation, which is suggestive of how exclusive the possession
of assets of the asset class EV is supposed to be.

Assume that A avails of political power that has been rightly obtained on the
basis of A's personal authority. Now it is clear that if A temporarily 
hands over power to B with an expectation of compensating returns, that 
constitutes a problem, equally much as it is considered when A hands over
power to B in order to allow B to profit from this.
Only if neither A nor B will profit from the transfer and it serves a good
purpose in general it will be allowed.%
}

I will assume that A intends to lend exclusive valuables V$_e$ (that is V$_e$ is of type EV)
to borrower B. Now it is reasonable that a transaction $\alpha$ of which this lending is
a constituent has no part that  matches either with BCT(EV,G) or with LCP(EV,G).
To see that this prohibition is  rational notice:

\begin{enumerate}
\item 
The fact that A is in the possession of V signifies that A
has acquired an important and exclusive position. Ownership of V$_e$, 
which is assumed to have been acquired in a legitimate fashion,
brings with it a responsibility for A to make use of that
ownership himself.%
\footnote{For instance A may buy real estate in exchange for
(a part of) V$_e$ and subsequently rent that to others, thus making sure that
others profit from A's actions.
}
\item
That temporal transferal of control over valuables V$_e$ by A to another agent B
is therefore unethical, irrespective of interest being paid, claimed,
or promised, in any direction.
\item
It is allowed, however, that A asks B to keep V$_e$ in store ready for A's later use
of it.
\end{enumerate}
I notice that, in the context of a theologically based endorsement of this prohibition, this
particular rule of prohibition has a number of intriguing properties:%
\footnote{In fact one finds the key properties that Subhani in \cite{Sub01a} claims to 
be in need of an explanation (which have been summarized as three questions in section  
\ref{questions} above).%
}
\begin{itemize}
\item 
This prohibition might be endorsed by some god (hereafter called God).
That endorsement can be promised by Gods authoritative priests or scholars.
Assuming that endorsement:
\item 
Violations of this prohibition are punished by God only because those who
own exclusive valuables are in power by definition and earthly
punishment cannot be adequately administered to them.

\item
The punishment promised on violation is very severe because it is essential that
the wealthy go at lengths to realize a just society by their own action. Only
by giving away their wealth the can escape from the significant obligations that go
with wealth.

\item 
An offense against this prohibition need not damage any one else except for
the fact that the owner of V$_e$ by failing to live up to his duties fails to contribute
to the best operation of society, a course of action that A owes to God.
As a consequence promised punishment need not be based on
third party damage.

\item 
Because it is a matter of principle the degree to which control is transferred,
concerning the use of V$_e$
as measured by the productive capacity which goes along with its
transferal does not influence the promised penalties. In other words any 
non-zero production of G results in an offense.

\item 
If money (say fiat money FM) is not classified as exclusive (EV) it is conceivable that
prohibition of CPT(EV,G) coexists with permission of CPT(FM,FM).
\end{itemize}

\section{Dialectical roots do not imply the prohibition of SAI}
I will now consider each of the dialectical roots for IP and one by one it will be argued that 
the root cannot be used to infer that SAI must be prohibited. Because these dialectical roots
cover all rational arguments for IP, it follows that no such rational argument exists in the 
specific case of SAI. This reasoning may be criticized for ignoring the possibility that a
combined use of different dialectical roots might lead to a convincing rejection of SAI. Combined use of two or more roots is only possible if these assume compatible 
perspectives on money. Lacking the ability to analyze this matter conclusively I must admit 
that the main conclusion if this paper, that the justification SAI prohibition necessarily
involves a chain of arguments from authority has been made plausible but has not been
secured with complete certainty.

The absence of an argument from a root case to SAI is understood as the implausibility of 
the presence of an analogy (see \cite{Mas98a}) between the root case and the description of
SAI.

\subsection{Judaic tradition}
In the Judaic tradition the lender is supposed to  have a stronger position than the 
borrower. This is entirely different from what SAI offers where the borrower  has the 
stronger position by far.  I refer for this matter to \cite{Gordon1982}, who suggests that loans from Jews to other Jews were considered consumptive credits.

\subsection{Aristotle}
A savings account can very well be used to store a reservation. The obtained
interest may be an unintended and even marginal side-effect. Prohibition of receiving
interest for an SAI cannot be based on the hypothesis of accumulation for its own sake,
conversely, the lender may accept a low interest rate because he sympathizes with how 
the borrower (the bank in the case of SAI) runs its business.

\subsection{Doubling debt scenario. }
The doubling debt scenario can be understood as being excessively harsh towards
a borrower incapable of returning the principal sum at the date of maturity. It represents 
usury in a morally objectionable form (taking as a part of its meaning the excessive 
imposition of a burden on the borrower's side).
There is no impact on the status of SAI, however, because in SAI the risk of default
for the borrower is close to zero because the borrower is supposed to possess ample
holdings available for liquidation when needed.

\subsection{Savings monopoly}
This root is tricky because although it may be thought of as a reason for the Church
to oppose positive interest rates it hardly qualifies as a justified argument. From an
Islamic point of view the monopoly that the Church may have intended to preserve for
itself lacks justification, and so do all other monopolies.

I conclude that it is a virtue rather than a flaw of SAI that it counteracts a savings 
monopoly, whoever may be profiting from it.

\subsection{Aquinas}
Aquinas' objections come quite close to making sense for SAI. It is no wonder that exactly 
his arguments have dominated the views of the Church concerning interest
for centuries. 

But if the bank offers an interest rate lower than what it usually earns when 
putting lenders' money into operation it is clear that no duplication needs to take place.

The double compensation objection is connected with several other and more 
recent objections: it might be considered a precursor of the circulation 
theory argument (\ref{cta} below), it might also be considered a rendering of the argument of strictly bifunctional money (\ref{sbfma} below).

Two aspects are underrepresented in the DC objection: the circulating nature of money and
the presence of borrower default risk. But there is something to it as well: if it is claimed that
the borrowing bank in the case of SAI will not default because it can sell (liquidate) its 
holdings when needed there is no guarantee that the money needed to buy those assets
exists in the system.

The DCO objection is somehow connected with the necessity of either economic growth or inflation in an economic system that makes use of interest based loans on a significant scale.
But it fails to explain why a categorical rejection of SAI is in place.

\subsection{Ex-sui creation}
Subhani's argument about ex-sui creation is ineffective to explain what might be wrong with
SAI provided the borrowing bank can produce the required interests by normal trade 
and investment  operations. The creation is not simply ex-sui, because the bank adds
its own unique expertise and its superior management competence.

The very reason that the bank offers an interest rate is based on its self-cconfidence that
it knows best how to turn the money into a profitable and useful investment.

\subsection{Strictly bifunctional money}\label{sbfma}
This way of arguing against SAI fails on the observation that no proof of concept for
I(MOE+SOV)  has been provided. 

In different words: only after it has been established that I(MOE+SOV) provides an 
adequate basis for a financial system, one may claim that its simplicity and its logical coherence per se constitute morally dependable grounds for IP.

\subsection{Circulation theory}\label{cta}
If the amounts saved are low the circulation theory argument does not
invalidate SAI as a reasonable mechanism. The objection of circulation theory seems
to depend on the assumption that a principal amount will always be returned. But that
assumption is unrealistic. Having that assumption out of the way interests can be
explained as an insurance against borrower default.

\subsection{Interest prohibition for ideal and fair money.}
If money is designed so that usage of borrowed money is rejected on moral grounds
interest is rejected as a consequence and it need not be independently forbidden.

\subsection{Personalized electronic money}
Equally, if debt cannot occur interest does not exist and for
that reason it cannot be forbidden.

This kind of argument has been used for the case of division by zero: who claims that
this is forbidden (not allowed), implicitly admits that it is possible. See \cite{BM09}

\subsection{Limited double compensation objection}
This mechanism seems to be more permissive than Islamic Finance is. It cannot be used 
to argue for a strict prohibition of interest in the case of SAI. On the contrary, this mechanism
can explain a legal amount due for borrowing money under the assumption that some fixed fraction of the money will be kept available by the borrower, who promises not to make use
of it for any form of expenditure. In this case the money serves as a MOE (means of exchange), a SOV (store of value), and a MOPP (manifestation of purchasing power).

\subsection{Credit money only}
This is close to the current situation. Even if many prefer other virtues from a borrower than
promised interest the fact that other criteria come into play does not invalidate interests
as a means for a bank to attract SAI's from potential customers. A categorical rejection of SAI cannot be inferred from this root.

\subsection{Prohibition of CP(EV,G)}
This root case is inapplicable for arguing against SAI, as it assumes that the lending 
party is the stronger one, which is not the case in SAI.

\section{Concluding remarks}
In~\cite{Ber11a} an attempt is made to formulate how logic can be helpful for decision 
making in an Islamic context. In principle it is conceivable that, after an extensive
period of development of so-called Real Islamic Logic (RIL) that apparatus can become helpful
to re-analyze the complete corpus of writings on interest prohibition and to come up
with a rationale for IP that has yet been missed so to speak. Analyzing this large and 
heterogenous body of texts is a daunting task, and it may be doable only with the
help of recent information technology (such as: ontologies, data mining, 
search engines for large volumes of text, automatic translation). 

There is no need to summarize the conclusions of the paper. More importantly its 
potential weaknesses can be stressed as  a form of risk assessment, as none of it 
has the rigor of mathematical proof. The following considerations might lead to 
dismissing the work done in part or in total:
\begin{itemize}
\item The possibility (and relevance) of making a 
sharp distinction between dialectical arguments and arguments from authority may
be disputed. 

For instance in \cite{Pal1994} a thorough analysis of IP is given which makes no
such distinction.
\item Dialectical root cases provide a way to obtain categorical injunctions against interest
for specific (proposed) financial products which cannot be found from any considerations
regarding negative effects of their use. This argument has been highlighted as the continuity
rule above in section \ref{CR}. If the continuity rule is rejected, the isolation and survey of dialectical root cases cannot play the role in analyzing IP that I have claimed it to play.
\item The suggestion that cut-elinination is possible for arguments from authority may be naive.
\item There may be an implicit bias in this account against the formation of new authorities
and for that reason against the inclusion of their views into an ASIL for IP. In particular the
trajectory an individual scholar moves along 
towards acquiring new authoritative status may involve
both a confrontation with the usage of arguments against specific financial products
based on root case similarity, as well as a an extensive
confrontation with negative practical consequences of ignoring IP in specific cases. 

Thus, although the continuity rule implies that expected disutility of positive interests cannot
justify their categorical prohibition, it is very well possible that observed disutility
constitutes a part of the background acquired by a new authority who 
bases his own judgement that IP is mandatory in part on his personal observations of
negative consequences of a system with interests. (This is similar to the arguments that may be used when imposing a strict ban on alcohol consumption.)
\end{itemize}

\bibliographystyle{plain}

\end{document}